\documentclass[12pt]{article}    

\usepackage{latexsym} 
\usepackage{amssymb}  
\usepackage{amsfonts} 
\usepackage{epsf}

\newcommand{\be}{\beta}
\newcommand{\ga}{\gamma}

\newcommand{\ra}{\rightarrow}

\newcommand{\dsp}{\displaystyle}

\newcommand\eqn[1]{(\ref{#1})}      
\newcommand\Eqn[1]{Eq.~(\ref{#1})}  
\newcommand{\e}{ {\rm e} }
\newcommand{\beq}{\begin{equation}}
\newcommand{\eeq}{\end{equation}}
\newcommand{\ba}{\begin{array}}
\newcommand{\bea}{\begin{eqnarray}}
\newcommand{\ea}{\end{array}}
\newcommand{\eea}{\end{eqnarray}}

\newcommand\comment[1]{ \hbox{[{\it Comment suppressed here.}\/]} }
\newcommand\hide[1]{}

\newcommand{\eg}{{e.g.}}

\newcommand{\skipover}[1]{}

\makeatletter 

\def\section{
\setcounter{equation}{0}        
\@startsection {section}{1}{\z@}{-3.5ex plus -1ex minus 
 -.2ex}{2.3ex plus .2ex}{\large\bf}}
\renewcommand{\theequation}{\arabic{section}.\arabic{equation}}

\def\subsection{\@startsection{subsection}{2}{\z@}{-3.25ex plus -1ex minus 
 -.2ex}{1.5ex plus .2ex}{\normalsize\bf}}

\def\subsubsection{\@startsection{subsubsection}{3}{\z@}{-3.25ex plus
 -1ex minus -.2ex}{1.5ex plus .2ex}{\normalsize}}

\makeatother   

\newsavebox{\eqlabel}

\makeatletter  
\newlength{\numblen}
\newsavebox{\eqnumb}
\def\@eqnnum{\savebox{\eqnumb}{\rm (\theequation)}%
\settowidth{\numblen}{\usebox{\eqnumb}}%
\makebox[\numblen][l]{\usebox{\eqnumb}~~~\usebox{\eqlabel}}}
\makeatother   

\newenvironment{equationwithlabel}[1]{ %
  \begin{equation}\label{#1} }{\end{equation}} 
\newcommand{\beql}[1]{\begin{equationwithlabel}{#1}}
\newcommand{\eeql}{\end{equationwithlabel}}

\begin{document}

\title{ \bf Imaginary chemical potential and finite fermion density 
on the lattice}

\newcommand{\ns}{\normalsize}

\author{
\begin{tabular}{c}
        Mark Alford, Anton Kapustin, Frank Wilczek \\
        \ns School of Natural Sciences \\ 
        \ns Institute for Advanced Study \\
        \ns Princeton, NJ 08540         
\end{tabular}
}

\newcommand{\preprintno}{
  \normalsize 
IASSNS-HEP-98/67
}

\date{\today \\ \preprintno}

\begin{titlepage}
\maketitle
\def\thepage{}          

\begin{abstract}

Standard lattice fermion algorithms run into the well-known sign
problem at real chemical potential.  In this paper we investigate the
possibility of using {\em imaginary} chemical potential, and argue
that it has advantages over other methods, particularly for probing
the physics at finite temperature as well as density.  As a feasibility
study, we present numerical results for the partition function of the
two-dimensional Hubbard model with imaginary chemical potential.

We also note that systems with a net imbalance of isospin  may be
simulated using a real chemical potential that couples to $I_3$
without suffering from the sign problem.

\end{abstract}


\end{titlepage}

\renewcommand{\thepage}{\arabic{page}}


\section{Introduction}

The behavior of fermions in the presence of a chemical potential is
relevant to condensed matter physics (Hubbard model away from half-filling)
and particle physics (high
quark density systems such as the early universe, neutron stars, and
heavy-ion collisions). Furthermore, a remarkably rich phase structure
has been conjectured for QCD at finite temperature and density
\cite{QuarkSupercond,PhaseDiag}.

The only reliable non-perturbative
approach to QCD is the numerical Monte-Carlo
evaluation of the functional integral using a lattice regulator.
Unfortunately, standard Monte-Carlo methods become inapplicable at
finite quark density, since in the presence of a real chemical
potential the measure is no longer positive.  One approach to
this problem is the ``Glasgow method'' \cite{Glasgow}, in which the
partition function is expanded in powers of $\e^{\be\mu}$, and the
coefficients are evaluated by Monte-Carlo, using an ensemble of
configurations weighted by the $\mu=0$ action.  Simulations using this
method have so far given unphysical results, namely, the lattice
starts to fill with baryons at a chemical potential well below the
expected value of one-third the baryon mass. It seems plausible that
this happens because the $\mu=0$ ensemble does not overlap
sufficiently with the finite-density states of interest, and so the
true effects of quark loops will only be seen at exponentially large
statistics \cite{Glasgow}.

In this paper we look at an alternative: evaluating the partition
function at {\em imaginary} chemical potential, for which the measure
remains positive, and standard Monte-Carlo methods apply.  The
canonical partition functions can then be obtained by a Fourier
transform \cite{dmst,ht}. Since the dominant source of errors is now the
Fourier transform rather than poor overlap of the measure, it seems
worthwhile to explore imaginary chemical potential as an alternative
to the Glasgow method.

An outline of the paper is as follows.
We give criteria that a theory should satisfy in
order for Monte-Carlo simulations at finite density to be feasible. 
We describe a toy model where even-odd effects effects become visible.
We find some interesting examples (\eg~QCD at finite isospin density)
where lattice simulations are possible. 
As a feasibility study we perform Monte-Carlo simulations for the
two-dimensional Hubbard model with imaginary chemical potential,
and find that it is indeed possible to obtain the canonical partition
functions at low particle number. At the rather high temperature
and low interaction strength that we study, we see no sign of 
electron pairing.

\section{Chemical potential and positivity of the measure}
\label{sec:gen}

Consider a generic system of fermions $\psi$ and bosons $\phi$,
where the fermion Lagrange density is $\bar\psi M(\phi) \psi$.
On integrating out the fermions, the partition function becomes
\beql{gen:Zdef}
Z=\int {\cal D}\phi\ \e^{-S_{bos}(\phi)} \det M(\phi).
\eeql
In order to perform Monte-Carlo simulations it is necessary that the
measure be nonnegative, so either we have to restrict ourselves to the
cases where $\det M\geqslant 0$, or to treat $\det M$ as an observable. The
latter option is usually not viable, as $\det M$ tends to be a
rapidly varying function of $\phi$. We will discuss it again at
the end of this section.

To guarantee that the measure is positive, we must generally have
an even number of flavors, for each of which $\det M$ is real
(but not necessarily positive).
One situation where $\det M$ is real is when there exists an invertible
operator $P$ such that
\beql{gen:dagger}
M^\dagger=P M P^{-1}.
\eeql
For a Wilson lattice fermion at zero chemical potential this relation holds,
with $P=\gamma_5$, so any even number of flavors can be simulated by
Monte-Carlo. With real chemical potential \eqn{gen:dagger} breaks
down, but with imaginary chemical potential it is valid, and again
simulations are possible for an even number of flavors.

There are more exotic situations where \eqn{gen:dagger} holds.
For example, consider two-flavor QCD with a finite density of isospin.
In this case $M$ has a block-diagonal structure
\beql{gen:L}
M(\mu)=\left( \begin{array}{cc} L(\mu) & 0       \\
                          0     & L(-\mu) 
              \end{array} \right),
\eeql
where $\mu$ is the chemical potential for the isospin, and $L(\mu)$ is
a Dirac operator for one flavor with chemical potential $\mu$.
$L(\mu)$ satisfies $L(\mu)^\dagger=\ga_5 L(-\mu)\ga_5 $, 
hence \Eqn{gen:dagger} is satisfied by setting 
\beq
P=\left( \begin{array}{cc} 0 & \ga_5 \\
                       \ga_5 & 0 
         \end{array} \right).
\eeq
Here $\det M(\mu)=|\det L(\mu)|^2\geqslant 0$.
More generally, consider QCD with $N_f$ flavors. It has a global vector-like
symmetry $G=U(1)_B\times SU(N_f)$, where $U(1)_B$ is the baryon number.
One may consider a nonzero chemical potential coupled to
any generator $T$ of $G$. Then Eq.~(\ref{gen:dagger}) is satisfied for
some choice of $P$ if and only if the nonzero eigenvalues of $T$ come in pairs
$\lambda,-\lambda$. Thus $\det M$ is real for QCD at nonzero isospin density,
but not for nonzero hypercharge density, or baryon number density. 

Another case where $\det M$ is real even in the presence of the
chemical potential is when there exists an invertible operator $Q$
such that
\beql{gen:conj}
M^*=QMQ^{-1}.
\eeql
Examples of this sort are afforded by models with four-fermion
interactions, like the Hubbard model and the Gross-Neveu model. There
$M$ is a real operator, so $\det M$ is real too. Other examples are
gauge theories with ``quarks'' in the real or pseudoreal
representation of the gauge group.  Thus $\det M$ is real for $SU(2)$
with quarks in the fundamental representation, or for $SU(3)$ with
quarks in the adjoint, even when the chemical potential is nonzero.

In some cases with $\det M$ real but not positive it appears that one
can perform simulations by treating the sign of $\det M$ as an
observable. The Hubbard model can be treated in this way far below
half-filling \cite{hubsim}.  This is also the case for the Gross-Neveu
model at nonzero chemical potential~\cite{KKW}. Ref.~\cite{KKW}
further argues that that $\det M$ is nonnegative for most of the
configurations, so one can simply replace $\det M$ with $|\det M|$.
In QCD the phase of the determinant contains important physical information,
but calculations have been performed without it \cite{Azcoiti}.


\section{Imaginary chemical potential}
\label{sec:im}

The fact that the fermion determinant for QCD is real in the presence
of an imaginary chemical potential $\mu=i\nu$ makes this
an attractive option for exploring finite quark density.
Simulations with an imaginary chemical potential are more or less
equivalent to simulating a canonical ensemble.  Indeed, the partition
function for imaginary chemical potential
\beq
Z(i\nu)={\rm Tr}\ \e^{-\beta \hat{H}} \e^{i\beta\nu \hat{N}},
\eeq
which is a periodic function of $\nu$ with period $2\pi/\beta$, is the
Fourier transform of the canonical partition function
\beql{im:fourier}
Z(N)=\frac{\beta}{2\pi} \int_0^{2\pi/\beta} d\nu Z(i\nu) \e^{-i\beta\nu N}.
\eeql
In principle, one can compute $Z(i\nu)$ on a lattice as a function of
$\nu$, and then use Eq.~(\ref{im:fourier}) to obtain the canonical
partition function.  In practice, this method can work only for low
enough $N$, because for large $N$ the integrand in
Eq.~(\ref{im:fourier}) is a rapidly oscillating function, and the
error of the numerical integration will grow exponentially with $N$.
The method fails completely in the thermodynamic limit
$N\ra\infty$. This need not discourage us, however, because in lattice
simulations one is always working in a finite and rather small
volume. The real question is how high we can push $N$ before
the numerical integration in Eq.~(\ref{im:fourier}) becomes undoable.  We will
consider the two-dimensional Hubbard model as a testing ground for this
approach. 
Related work has been performed in Ref.~\cite{dmst},
but without using the freedom to simulate at any $\nu$ (see below).

Having a positive measure is not the end of the story.
In practice we want to be able to use importance sampling 
to calculate $Z(i\nu)$ with reasonable accuracy. To this
end we rewrite $Z(i\nu)$ in the following form:
\beql{im:Znu}
{Z(i\nu)\over Z(i\nu_0)} =\int {\cal D}\phi\ \e^{-S_{bos}} 
\det M(i\nu_0) \, \frac{\det M(i\nu)}{\det M(i\nu_0)}.
\eeql
Now we treat the ratio of determinants as an observable, and the rest
as the measure. A natural worry is that the ratio of determinants
could be a rapidly varying function of the bosonic fields 
$\phi$, which would make
Monte-Carlo simulations unfeasible.  This is what
happens for {\em real} chemical potential. However, with an imaginary
chemical potential this does not occur. The absolute value of the
observable could still be a rapidly varying function. This
has to be decided on a case-by-case basis. We will see that
for the two-dimensional Hubbard model the ratio of determinants is a
slowly varying function of $\phi$, hence $Z(i\nu)$ is computable.

Using \eqn{im:Znu} we can calculate the partition function for a range
of $\nu$ around a reference value $\nu_0$. It is clear that we can
cover the range $\nu = 0\ldots 2\pi/\beta$ with a set of ``patches''
each centered on a different $\nu_0$. We can use as many patches as
are required, so the measure will always overlap arbitrarily well with
the observables. 

Some qualitative features of the system can be inferred from the
knowledge of $Z(i\nu)$ alone, without performing the Fourier
transform.  For example, consider a model of interacting fermions on a
lattice (the example of the Hubbard model will be discussed further below).
At some filling fraction
the system may undergo a phase transition to a BCS superconducting
phase.  In that phase, the system will be populated with Cooper pairs,
so the partition function will be dominated by sectors in which the
charge is a multiple of 2. This will be clearly visible in $Z(i\nu)$,
which will not only be perodic with period $2\pi/\beta$, but 
acquire a significant subharmonic at period $2\pi/(2\beta)$.  This
would be a signal that for some range of densities the energy of the
system has a minimum when the number of electrons is a multiple of 2.

This can be illustrated by a simple toy model containing a
fermion with mass $M_f$ and charge 1, and a boson with mass $M_b$ and
charge 2.  If we make $M_b$ less than $M_f$ then, assuming some very
weak charge-conserving interactions that establish equilibrium, 
states of even charge will be favored.

The free energy is the sum of the fermion and boson contributions,  
\beql{im:Ftot}
F(\mu,T) =  F_{\rm fermion}(M_f,\mu,T) + F_{\rm boson}(M_b,2\mu,T),
\eeql
where $F_{\rm fermion}$ and $F_{\rm boson}$ 
are the free energies of free fermions and bosons, respectively,
\beql{im:Fferm}
\ba{rcl}
F_{\begin{array}{c}\hbox{\rm\scriptsize fermion} \\[-1ex]
   \hbox{\rm\scriptsize boson} \end{array}}
(M,\mu,T) 
&=& \dsp \int {d^3p\over (2\pi)^3} \pm \ln\Bigl(
1 \pm 2\cosh(\be\mu)\e^{-\be E(p)} + \e^{-2\be E(p)} \Bigr), \\[2ex]
E(p)^2 &=& p^2 + M^2.
\ea
\eeql

\begin{figure}[Htb]
\begin{center}
\parbox{3in}{
\epsfxsize=2.7truein
\epsfbox{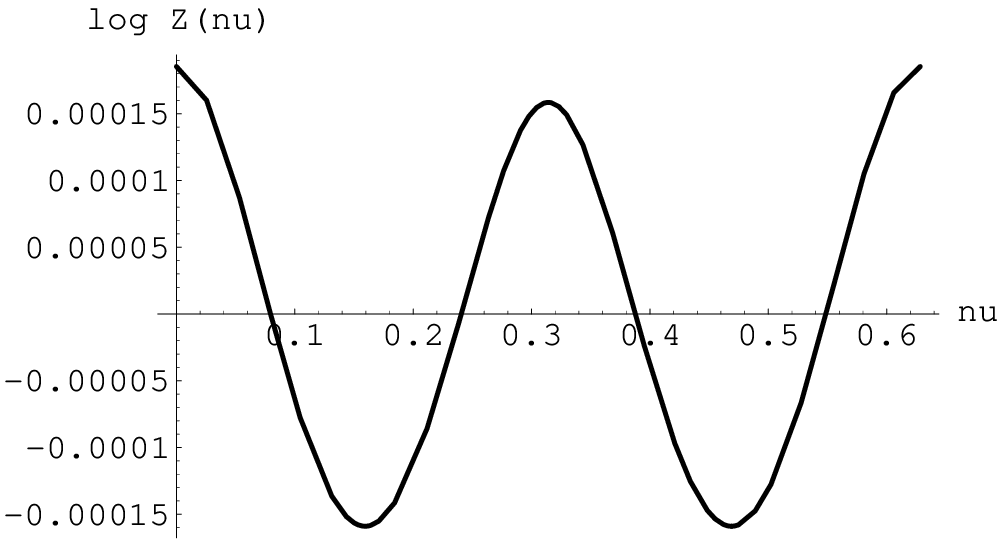}
 }
\hspace{0.2in}
\parbox{3in}{
\epsfxsize=2.7truein
\epsfbox{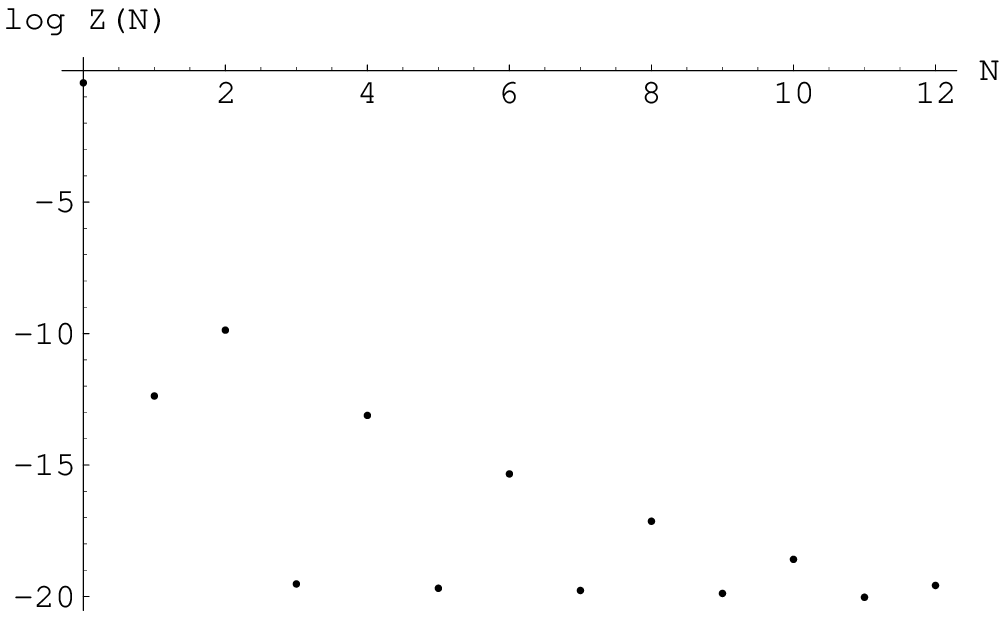}
 }
\end{center}
\vspace{-4ex}
\caption{ 
Free energy as a function of imaginary
chemical potential $\nu$ (left) and log of the canonical partition function
as a function of particle number (right) for our illustrative toy model
\protect\eqn{im:Ftot}, with $\be M_b = 1$ and $\be M_f = 5$.
}
\label{fig:lnZnu}
\end{figure}

The pairing of fermions into bosons is clearly visible
in $Z(N)$ (Fig.~\ref{fig:lnZnu}), and also directly in $Z(i\nu)$ which
is not only perodic with period
$2\pi/\beta$, but also approximately periodic with a smaller period
$2\pi/(2\beta)$.  This is a signal that
the energy of a system has a minimum
when the number of particles is a multiple of 2. 
However, to infer the existence of an ``unpairing'' transition at high
chemical potential, a visual inspection of the the plot of $Z(i\nu)$
would not suffice: this information is encoded in the
high-frequency behaviour of the Fourier transform of $Z(i\nu)$.

\section{Hubbard model with imaginary chemical potential}
\label{sec:hub}

At densities away from half-filling, the single-flavor Hubbard model
has a real but not necessarily positive fermionic
determinant \cite{Creutz}. Like QCD, the measure becomes positive at imaginary
chemical potential.  Since this model is of physical interest
and also much less
computationally demanding than QCD, it is interesting to use it to
study the feasibility of performing simulations with imaginary
chemical potential.  In fact, the model is so simple that for this
initial investigation we were able to dispense with the usual hybrid
Monte-Carlo algorithm \cite{Creutz} for evaluating the fermion
determinant, and perform the whole calculation with the computer
mathematics tool ``Mathematica'', using its ``Det'' function to
calculate the fermionic determinants.

\begin{figure}[Htb]
\begin{center}
\epsfxsize=4in
\epsffile{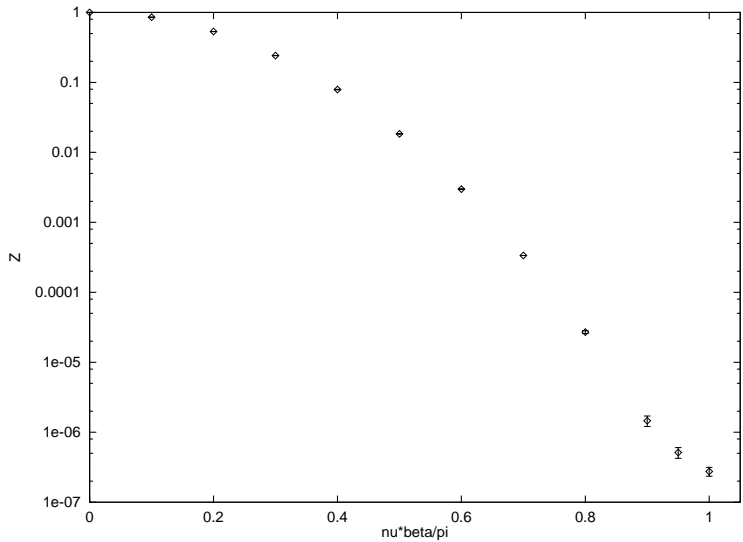}
\end{center}
\caption{ 
Partition function for 2D Hubbard model as
a function of imaginary
chemical potential $\nu$, in units of $\pi/\be$.
This is on a $4^2\times 10$ lattice, with $\be=1.5$, 
hopping term $K=1$, and interaction strength $U=0.1$,
following the conventions of Creutz \cite{Creutz}.
}
\label{fig:lnZhub}
\end{figure}

Using the formulation described above (see \eqn{im:Znu}), we
calculated the partition function as a function of imaginary chemical
potential $\nu$.  We used a $4^2\times 10$ lattice with inverse
temperature $\be=1.5$.  (Lower statistics were also gathered on
$4^2\times 20$, to check that temporal discretization errors were
under control.)  The results are given in Fig.~\ref{fig:lnZhub}.  By
particle-hole symmetry, $Z(i\nu) = Z(-i\nu)$, and $Z$ has period
$2\pi/\beta$, so we only plot $\nu=0$ to $\pi/\beta$.

Three ``patches'' were used (see Sect.~\ref{sec:im}),
centered at $\beta\nu_0 = 0, \pi/2$, and $\pi$.  At the temperature we
study, the error in $Z(i\nu)/Z(i\nu_0)$ rises rapidly with
$|\nu-\nu_0|$, so it is crucial to use multiple patches to keep the
statistical errors in $Z(i\nu)$ under control. In turn, via the
Fourier transform, this controls the errors in $Z(N)$ for $N>0$.  In
contrast, Ref.~\cite{dmst} only used $\nu_0=0$, which is adequate only
for the small volume, low temperature, and low particle number
($N\leqslant 2$) they studied.

\begin{figure}[Htb]
\begin{center}
\epsfxsize=4in
\epsffile{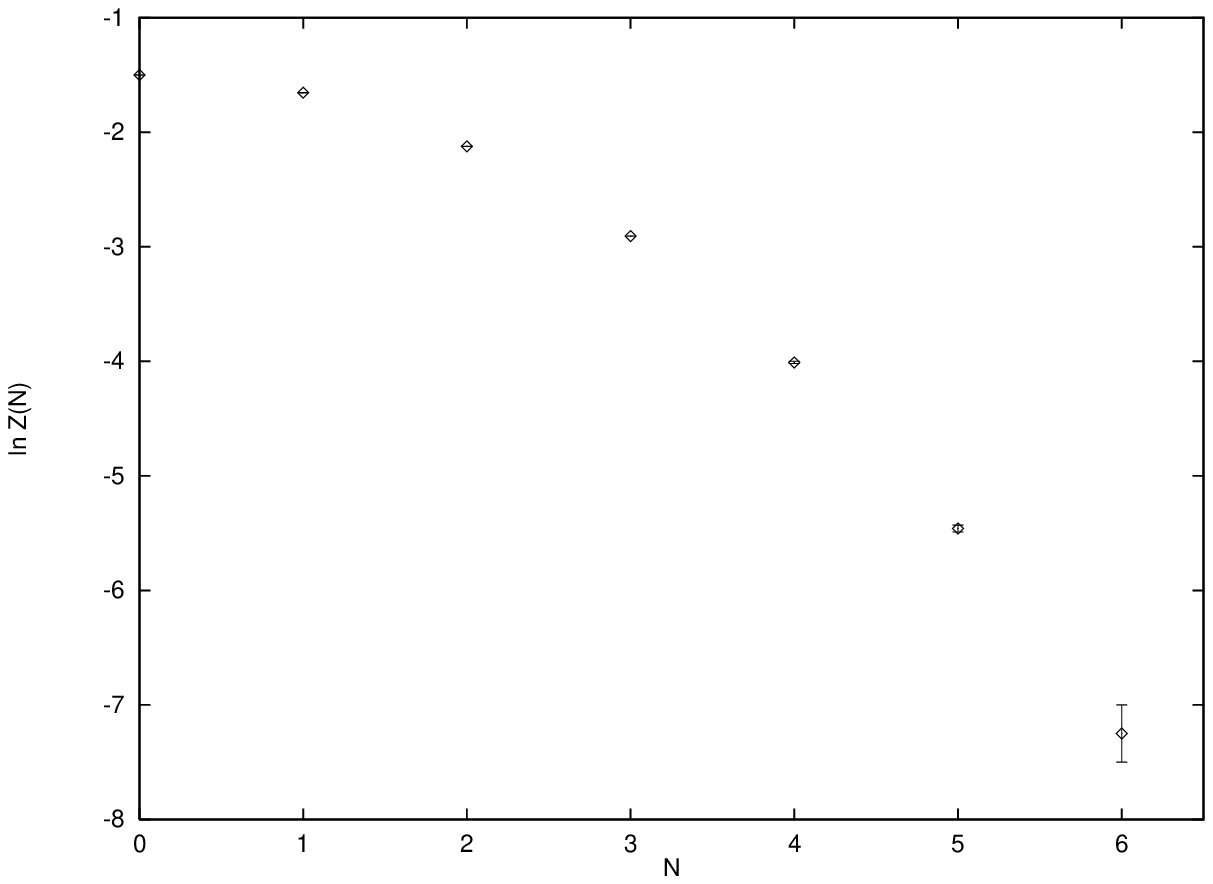}
\end{center}
\caption{ 
Log of canonical partition function $Z_N$ for 2D Hubbard model,
obtained by Fourier transform of $Z(i\nu)$ (Fig.~\protect\ref{fig:lnZhub}).
}
\label{fig:ZNhub}
\end{figure}

We then fitted $Z(i\nu)$ to various ansatze, and Fourier transformed them
to obtain the canonical partition function $Z_N$ (Fig.~\ref{fig:ZNhub}).
Even using our very inefficient updating algorithm, we are easily able to
to get accurate results up to $N=6$. The error bars reflect the 
statistical errors in the $Z(i\nu)$ data.
We used fit functions of the form
$Z(i\nu)=\exp(-a\nu^2)\times {\rm spline}$.

It has been suggested 
that at some filling fraction
the 2D Hubbard model may exhibit superconductivity, through pairing of
the electrons to form Cooper pairs which then condense.  Along the
lines described in section \ref{sec:im}, we would expect such pairing
to manifest itself as an even-odd periodicity of $Z_N$, leading to a
characteristic extra bump in $Z(i\nu)$. At the temperature and
coupling that we studied, we see no such evidence of
pairing. Obviously it would be very interesting to explore a wider
range of parameters.  This method might also be suitable for exploring
the metal-insulator transition near half-filling, where the sign
problem becomes particularly virulent \cite{hubsim}.
As noted in Ref.~\cite{dmst}, however, it will become much harder
to extract the $Z_N$ at low temperatures, where  the $N=0$ contribution
dominates $Z(i\nu)$.

\section{Conclusions}

Let us conclude with a few remarks concerning the possible utility of
imaginary chemical potential in QCD.  An imaginary chemical potential
does not systematically bias the ensemble to large density, so that to
pick out the effect of states of non-zero baryon number density one
must rely on fluctuations (which, when they occur, are appropriately
weighted by the action).  These fluctuations will occur most readily
when the temperature is high, and the gap in the baryon number channel
is small. And even then one can only realistically hope, on the small
lattices likely to be practical, to fluctuate to a few baryons.  So a
reasonable procedure would seem to be to start with a high temperature
and work down, looking for qualitative changes as a function of
temperature.  In this way, one could realistically hope to 
use the methods described in this paper to study how
properties of the quark-gluon plasma are affected by a net quark
density.  
One could also study the deconfinement crossover, near which
the baryons become light, and in particular locate the critical point
in the $T-\mu$ plane predicted for two-flavor QCD
\cite{PhaseDiag,tricrit}.  All these phenomena are of immediate
interest, since they will be explored in the next generation of heavy
ion collision experiments.

Another possibility is to work with large numbers of quark species,
close to 16, so that the theory is perturbative.  Then there is no
mass gap to baryons, so fluctuations are cheap, and also the
contribution of interest, due to the quarks, is not swamped by gluons.
In this case the cancellations may not be so bad even at real chemical
potential, and the imaginary chemical potential approach should also
work better, since the fluctuations of interest will occur frequently.

\vspace{3ex}
{\samepage 
\begin{center} Acknowledgements \end{center}
\nopagebreak
The research of MGA, AK and FW is supported by DOE grant DE-FG02-90ER40542.
MGA is additionally supported by the generosity of Frank and Peggy Taplin.
\par
}

\end{document}